\documentclass[twocolumn,letter]{jpsj3}
\usepackage{txfonts}

\title{High-Field Magnetization of CeRu$_2$Al$_{10}$}

\author{Akihiro \textsc{Kondo}$^{1}$\thanks{E-mail address: kondo@issp.u-tokyo.ac.jp}, Junfeng \textsc{Wang}$^{1, 2}$, Koichi \textsc{Kindo}$^{1}$, 
Tomoaki \textsc{Takesaka}$^{3}$, Yukihiro \textsc{Kawamura}$^{3}$, \\
Takashi \textsc{Nishioka}$^{3}$, Daiki \textsc{Tanaka}$^{4}$, Hiroshi \textsc{Tanida}$^{4}$ and Masafumi \textsc{Sera}$^{4}$}

\inst{$^{1}$Institute for Solid State Physics, University of Tokyo, Kashiwa, Chiba 277-8581, \\
$^{2}$Wuhan High Magnetic Field Center, Huazhong University of Science and Technology, Wuhan 430074, China, \\
$^{3}$Graduate school of Integrated Arts and Science, Kochi University, Kochi 780-8520, \\
$^{4}$Department of Quantum Matter, AdSM, Hiroshima University, Higashi-hiroshima 739-8530}

\abst{The magnetization measurements of Ce$_{x}$La$_{1-x}$Ru$_2$Al$_{10}$ ($x$ = 1, 0.75) under the high magnetic field were performed in order to obtain the information for the long-range order (LRO) in CeRu$_2$Al$_{10}$. 
We successfully obtained the magnetic phase diagram of these two compounds for the applied magnetic field along the $a$-axis which is the magnetization easy axis, and found that the LRO for $x$ = 1 disappears at $\sim50$ T which is the critical field to the paramagnetic phase. 
For $x$ = 0.75, the critical magnetic field decreases to $\sim37$ T by La substitution. 
The magnetic phase diagram and magnetization curve are qualitatively consistent with the recent Hanzawa's mean field calculation results obtained by assuming the dimer of Ce ions whose crystalline electric field ground state has a large magnetic anisotropy. 
These results support the singlet pair formation scenario recently proposed by Tanida $et\ al$..  
We also pointed out the possibility of the appearance of the field-induced magnetic phase between $\sim40$ T and $\sim50$ T for $x$ = 1.}

\kword{CeRu$_2$Al$_{10}$, spin gap, singlet ground state, Kondo semiconductor, pulsed high magnetic field}

\begin{document}
\maketitle

Recently, new ternary compounds Ce$T_2$Al$_{10}$ ($T$ = Fe, Ru, Os), with orthorhombic YbFe$_2$Al$_{10}$-type structure, have attracted much attention because of the unusual physical properties.\cite{st1, st2, Muro, Stry, Take, Nishi, Matsu, Tany1}
In particular, CeRu$_2$Al$_{10}$ is of great interest due to the novel phase transition at $T_0$ = 27 K. 
CeRu$_2$Al$_{10}$ exhibits the following characteristic features.\cite{Nishi, Matsu, Tany1} 
The specific heat, $C$ shows a clear peak accompanied with the $\lambda$-type anomaly. 
The magnetic entropy obtained from the specific heat measurements is $\sim$0.65 Rln 2 and Rln 2 at $T_0$ and $\sim100$ K, respectively.
This suggests that the crystalline electric field (CEF) ground state is doublet. 
The electrical resistivity, $\rho$ monotonically increases with decreasing temperature and a steep increase just below $T_0$, and rapidly decreases after showing a peak.
The magnetic susceptibility, $\chi$ shows a large anisotropy.
The magnetization easy axis is along the $a$-axis. 
In the case of the magnetic field ($H$) parallel to the $a$-axis, the Curie-like behavior can be clearly observed. 
Although it is important to know the CEF level scheme and the CEF ground state wave function, it is not known at present. 
Below $T_0$, $\chi$ shows a decrease along all the crystal axis. 
The results of $C$, $\rho$ and $\chi$ are well described  by the thermally activated form below $T_0$. 
This indicates the opening of a spin gap and/or a charge gap on the Fermi surface below $T_0$. 
The magnitude of gap is estimated to be $\Delta\sim100$ K. 

At present, there are several candidates for the origin of a long-range order (LRO) below $T_0$. 
At first, the possibility of the antiferro (AF) magnetic ordering was reported.\cite{Stry} 
However, the AF magnetic order in CeRu$_2$Al$_{10}$ was ruled out by the NQR measurement.\cite{Matsu} 
As another candidate of the LRO, the charge density wave (CDW) transition is suggested by the fact that some experimental results indicate the existence of gap.\cite{Nishi}  
Matsumura $et\ al$. proposed the nonmagnetic structural transition from the nonexistence of the AF mgnetic order by the NQR results.\cite{Matsu} 
On the other hand, Tanida $et\ al$. proposed that the phase transition originates from the singlet pair formation between Ce ions.\cite{Tany1} 
In La substituted system Ce$_{x}$La$_{1-x}$Ru$_2$Al$_{10}$, $T_0$ continuously decreases and the peak of $C$ becomes small and broad with decreasing $x$. 
$T_0$ can be recognized at $x\sim0.45$. 
Furthermore, $T_0$ slightly decreases with increasing $H$ up to 15 T. 
The thermal expansion experiments indicate that the crystal volume decreases below $T_0$. 
These results strongly suggest that the LRO has magnetic nature and the singlet ground state is constructed below $T_0$. 
Thus, CeRu$_2$Al$_{10}$ may be the first example showing the LRO with a singlet ground state among the rare-earth compounds. 
However, even if this is the case, several problems remain in this scenario, such as the origin of a large magnitude of the magnetic susceptibility below $T_0$ and the existence or the nonexistence of the low dimensionality in this orthorhombic structure. 
Therefore, the nature and mechanism of the LRO in CeRu$_2$Al$_{10}$ have not yet been clarified. 
Being different from the singlet ground state in 3$d$ compounds with $S$ = 1/2 where an ideal Heisenberg model is applicable, the orbital degrees of freedom should play some role in a singlet ground state in the rare-earth compounds. 
From the intensive studies on CeRu$_2$Al$_{10}$, new physics on the singlet ground state not seen in 3$d$ compounds will be produced. 

If the LRO with a singlet ground state is realized below $T_0$,  the existence of the critical field to a paramagnetic region is expected.  It is important to know the magnitude of the critical field from which the information on the magnitude of a spin gap is obtained.  It is also interesting to know the critical fields along three crystal axes in this compound with a large magnetic anisotropy which induced the information on the role of the orbital degrees of freedom in a spin gap in rare-earth compounds. 
The magnetic phase diagram obtained from the thermal expansion\cite{Tany1} suggests that the critical field from the LRO phase to the paramagnetic phase is expected to tens of tesla. 
In addition, the magnetization process shows no anomaly for the applied magnetic field along the $a$-axis up to 5 T.\cite{Matsu}  
Therefore, it is necessary to apply the higher magnetic field up to tens of tesla in order to clarify the magnetic properties of the LRO. 
In the present paper, we performed the magnetization measurements of Ce$_{x}$La$_{1-x}$Ru$_2$Al$_{10}$ ($x$ = 1, 0.75) under the high magnetic field in order to obtain the information for the LRO in CeRu$_2$Al$_{10}$.

The single crystals of Ce$_{x}$La$_{1-x}$Ru$_2$Al$_{10}$ ($x$ = 1, 0.75) used in the present study were prepared by the Al self-flux method.
Pulsed magnetic fields up to 55 T were generated with a duration time of 36 ms, using non-destructive magnets. 
Magnetization was measured by the induction method using a standard pick-up coil 
in magnetic fields along the $a$-axis which is the magnetization easy axis.  
The absolute value was calibrated by comparing the data with the magnetization below 7 T measured using MPMS (Quantum Design). 

\begin{figure}[tb]
\begin{center}
\includegraphics[width=0.85\linewidth]{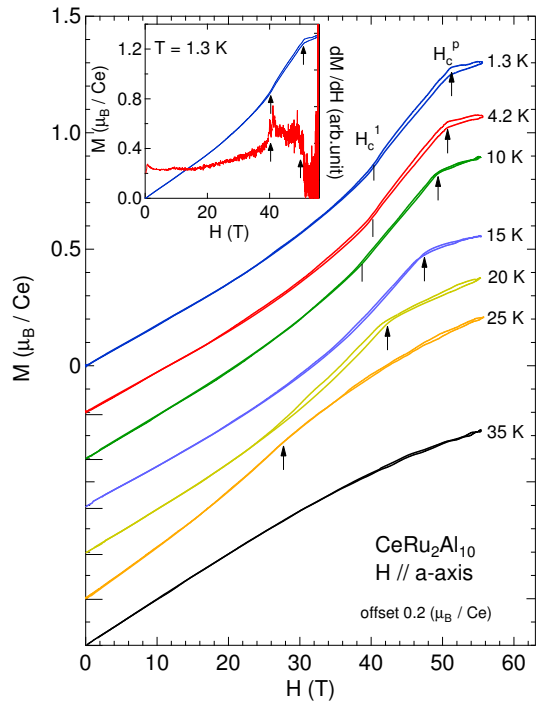}
\caption{Magnetic field dependence of the magnetization of CeRu$_2$Al$_{10}$ under various temperatures for $H\parallel$ a-axis. The origin of the vertical axis of each curve in magnetic field is shifted so as to see it easily.  The inset shows the magnetization curve and  derivative d$M$/d$H$ at $T$ = 1.3 K .}
\label{Fig1}
\end{center}
\end{figure}

\begin{figure}[tb]
\begin{center}
\includegraphics[width=.85\linewidth]{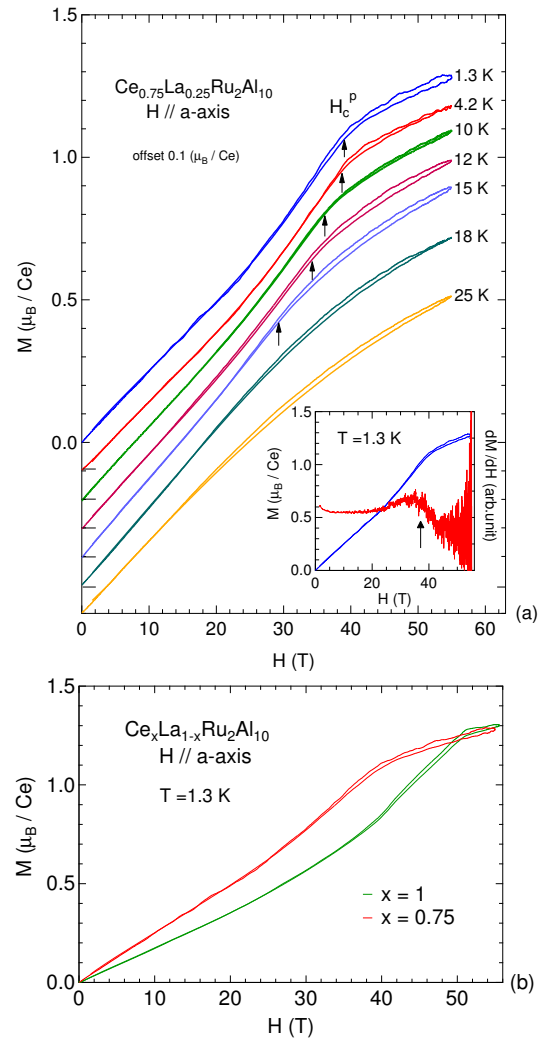}
\caption{(a) Magnetization curve of Ce$_{0.75}$La$_{0.25}$Ru$_2$Al$_{10}$ under various temperatures for $H\parallel$ a-axis. The origin of the vertical axis of each curve in magnetic field is shifted so as to see it easily.  The inset shows the magnetization curve and  derivative d$M$/d$H$ at $T$ = 1.3 K . (b) Magnetization curve of Ce$_{x}$La$_{1-x}$Ru$_2$Al$_{10}$ ($x$ = 1, 0.75) at $T$ = 1.3 K. }
\label{Fig3}
\end{center}
\end{figure}

\begin{figure}[tb]
\begin{center}
\includegraphics[width=.85\linewidth]{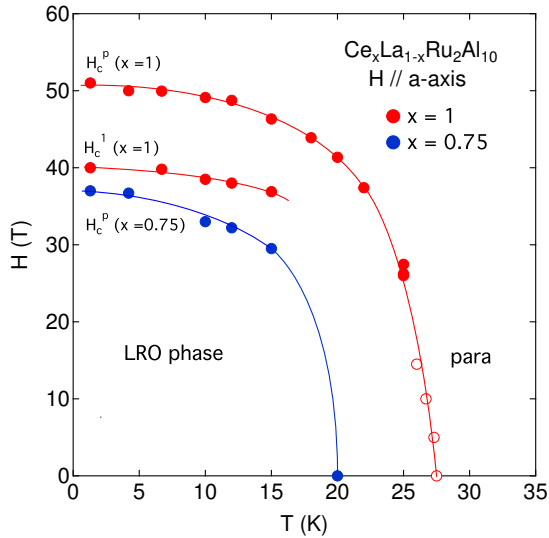}
\caption{Magnetic phase diagram of Ce$_{x}$La$_{1-x}$Ru$_2$Al$_{10}$ ($x$ = 1, 0.75) for $H\parallel$ $a$-axis. The data (red open circle) for $x$ = 1 below 15 T are cited from ref. 8.}
\label{Fig5}
\end{center}
\end{figure}

Figure 1 shows the $H$ dependence of the magnetization ($M$) of CeRu$_2$Al$_{10}$ for $H\parallel$ $a$-axis.  
The inset shows the magnetization curve and  derivative d$M$/d$H$ at 1.3 K .
At 1.3 K, $M$ shows the $H$-linear increase with a small slope of $M$ up to $\sim 30$ T. 
This is probably due to the Van Vleck contribution from the excited state, which will be discussed later. 
With further increase of $H$, $M$ shows a kink at $\sim40$ T and a downward bending at $\sim50$ T. 
Hereafter, the anomalies at $\sim40$ T and $\sim50$ T are called as $H_c^1$ and $H_c^p$, respectively. 
$H_c^p$ is the critical field from the LRO to the paramagnetic region.  
The present result of $H_c^p\sim50$ T is consistent with the magnitude of a gap $\Delta \sim100$ K below $T_0$ estimated from various properties in the sense that the order of the magnitude of $H_c^p$, $T_0$ and $\Delta$.\cite{Nishi, Matsu}   
As there exists large magnetic anisotropy in CeRu$_2$Al$_{10}$, $H_c^p$ is expected to depend on the applied field direction, which may induce the information on the spin gap in rare-earth compounds.  
These anomalies are clearly recognized by d$M$/d$H$, as shown in inset of Fig. 1. 
As temperature increases, $H_c^p$ shifts to lower magnetic field. 
In contrast, $H_c^1$ becomes unclear with increasing temperature and is barely recognized above 15 K.   
One of the  possible mechanisms for the intermediate phase between $H_c^1$ and $H_c^p$ is due to the field-induced transition as proposed by Tachiki and Yamada\cite{Tachi}, as we will discuss later.

Figure 2 (a) shows the magnetization process of Ce$_{0.75}$La$_{0.25}$Ru$_2$Al$_{10}$ for $H\parallel$ $a$-axis. 
The inset shows the magnetization curve and  derivative d$M$/d$H$ at $T$ = 1.3 K. 
$M$ at 1.3 K increases linearly with increasing $H$ as is observed for $x$ = 1, and shows an anomaly at $\sim37$ T.  
This anomaly corresponds to $H_c^p$ for $x$ = 1. 
With increasing temperature, $H_c^p$ shifts to lower magnetic field, and is confirmed to be $\sim$28 T at 15 K from the d$M$/d$H$-$H$ curve. 
On the other hand, the anomaly at $H_c^1$ for $x$ = 1 becomes unclear by La substitution, as shown in inset of Fig. 2 (a). 
Above $H_c^p$, $M$ monotonically increases as the magnetic field increases. 
The value of $M$ at 55 T for 1.3 K is $\sim$1.3$\mu_{\rm B}/$Ce, which is  almost the same as that for $x$ = 1, as shown in Fig. 2 (b).

Figure 3 shows the magnetic phase diagrams of Ce$_{x}$La$_{1-x}$Ru$_2$Al$_{10}$ ($x$ = 1, 0.75) for $H\parallel$ $a$-axis obtained by the present results. 
The data for $x$ = 1 below 15 T are cited from ref. 8.
For $x$ = 1, the LRO at $H$ = 0 continues to exist up to 40 T. 
With increasing $H$, $T_0$ shifts to lower temperatures. 
The critical magnetic field to the paramagnetic phase at 1.3 K for $x$ = 1, $H_c^p$, is $\sim50$ T. 
By La 25\% substitution, $H_c^p$ decreases down to $\sim37$ T. 
We note that the $x$ dependence of $H_c^p$ seems to be similar to that of $T_0$; These values are reduced by $\sim$25\% owing to the La 25\% substitution. 
These facts suggest that the origin of the LRO is magnetic. 
As described above, it is possible that the anomaly at $H_c^1$ is due to the jump just below the saturation of $M$.  
Recently, Tanida $et\  al.$ measured the magnetization process and the magnetoresistance of CeRu$_2$Al$_{10}$ for $H\parallel c$-axis  and  found the phase boundary or cross over at $H^*\sim$ 4 T below $T_0$ characterized by the metamagnetic like magnetization process.\cite{Tany3} 
They found that $H^*$ increases rapidly by rotating the magnetic field from the $c$-axis towards the $a$-axis above $\sim$ 60 degrees. 
This result implies that the boundary for $H\parallel a$-axis at $\sim40$ T is the same kind of boundary as that for $H\parallel c$-axis. 
In the $M$-$H$ curve, it is difficult to confirm $H_c^1$ above 15 K because the magnitude of the slope, d$H_c^1$/d$T$, becomes larger above $\sim$15 K.
It may be possible to observe $H_c^1$ in the temperature dependence of $M$ under the high magnetic field. 
We suspect that $H_c^p$ and $H_c^1$ coincide with each other at $H\sim$ 20 T because only one phase transition temperature, $T_0$ was observed below $H$ = 14.5 T.\cite{Tany1}

The important results obtained by the present study are that the LRO in CeRu$_2$Al$_{10}$ has the magnetic nature and the magnetic phase diagram for $H\parallel a$ is obtained. 
At present, the most probable candidate for the LRO is the singlet pair formation between Ce ions proposed by Tanida $et\ al$. \cite{Tany1} 
Here, from this standpoint, we discuss the field induced phase at high magnetic fields.
In the most simple case of the singlet ground state formed by the two magnetic ions with $S$ = 1/2, the metamagnetic transition should be observed at the critical field where level crossing occurs between the singlet ground state and one of the excited triplet. 
In this case the Zeeman energy loss of the singlet state becomes larger with increasing magnetic field up to the critical field.  
In a real system, in order to release such a large Zeeman energy loss, the intermediate magnetic phase appears at higher magnetic fields below the critical field to the paramagnetic phase. 
In the spin-Peierls system, it is well known that the magnetic phase, called $M$ phase, exists in the intermediate magnetic fields. 
For example, in a typical spin-Peierls compound CuGeO$_3$, the critical magnetic field to $M$ phase and to the paramagnetic phase is 12 T and 253 T, respectively.\cite{CuGe1, CuGe2} 
These phase diagrams are obviously different from that of CeRu$_2$Al$_{10}$. 
Thus, CeRu$_2$Al$_{10}$ is not categorized as a conventional spin-Peierls system observed in 3$d$-electron system. 
On the other hand, Tachiki $et\ al$. investigated theoretically the system where the inter-pair interaction between two spin singlet pairs was taken into account in addition to the intra-pair interaction, and showed that the spin-ordered phase appears in the vicinity of the point of level crossing. \cite{Tachi}
The $M$-$H$ curve obtained from the calculation shows a kink at the critical magnetic field from the singlet ground state to spin-ordered phase, which is similar to that of CeRu$_2$Al$_{10}$ at $\sim40$ T. 
It is, however, quite difficult to apply their model immediately to CeRu$_2$Al$_{10}$ because the orbital degrees of freedom are not considered in their model. 
It is desirable to perform the calculation involving the orbital degrees of freedom in the future.

Quite recently, Hanzawa investigated theoretically the singlet ground state formed by the dimer of the two Ce ions with a large magnetic anisotropy in the CEF doublet ground state\cite{Han}. 
He assumed that two Ce ions construct a singlet ground state via RKKY interaction, and performed the mean field calculation. 
In this model, the main part of the CEF ground state is $|\pm3/2\rangle$ and the component of $|\pm5/2\rangle$ is contained slightly. 
The weight of $|\pm5/2\rangle$ and $|\pm3/2\rangle$ of the CEF ground state wave function is $a$ : $\sqrt{1-a^2}$, respectively. 
Here, $a$ is a real number. 
This assumption is consistent with the present experimental results of $M$-$H$ curve. 
He could reproduce the existence of the large Van Vleck term below $T_0$ together with its large anisotropy. 
The $M$-$H$ curve was also calculated. 
In the case of $a\neq0$, the calculated $M$-$H$ curve shows a linear increase due to the Van Vleck contribution from excited state, and shows a metamagnetic transition to the paramagnetic phase.  
When $a^2=2/10$, the critical field at $T$ = 0 is 54.2 T. 
This critical field is consistent with the experimental result where $H_c^p$ is $\sim50$ T. 
In addition, the slope of the $M$-$H$ curve below $\sim30$ T is almost similar to that of the present experimental data. 
These similarities between the calculated and experimental results imply that the singlet ground state formed by the dimer of Ce ions is realized in CeRu$_2$Al$_{10}$ below $T_0$. 
However, there exists a difference between the calculated results and experimental one at high magnetic fields.  
While the calculated $M$-$H$ curve shows a metamagnetic transition at critical field to the paramagnetic phase,  the experiments show two anomalies at $H_c^1$ and $H_c^p$. 
These discrepancies indicate that the singlet ground state realized in CeRu$_2$Al$_{10}$ is unconventional state.

In summary, we performed the magnetization measurements of Ce$_{x}$La$_{1-x}$Ru$_2$Al$_{10}$ ($x$ = 1, 0.75) under the high magnetic field. 
We found that the LRO for $x$ = 1 disappears at $\sim50$ T for $H\parallel a$-axis. 
For $x$ = 0.75, the critical magnetic field to the paramagnetic phase decreases to $\sim37$ T by La 25\% substitution. 
The $H$-$T$ diagram and $M$-$H$ curve are qualitatively consistent with those of the mean field calculation results by Hanzawa. 
These results support the recently proposed singlet ground state scenario by Tanida $et\ al$..  
We also pointed out the possibility of field-induced magnetic phase. 
The origin of the field-induced phase is not clear at present.  
It is important  to know $H_c^p$ for $H\parallel c$ and $b$-axis from which the information on the magnitude of the anisotropy of a spin gap could be induced.  
We have a plan to measure the magnetization for $H\parallel c$ and $b$-axis at high magnetic fields in the near future. 
Further studies are necessary to clarify the nature of the LRO in CeRu$_2$Al$_{10}$. 

In the course of the present study, a small internal field was observed below $T_0$ of CeRu$_2$Al$_{10}$ by $\mu$SR\cite{mSR} and a spin gap with the excitation energy of $\sim8$ meV which grew up below $T_0$ was directly observed by the inelastic neutron scattering\cite{ND}. 
The magnetic excitation energy of $\sim8$ meV is consistent with the present result of $H_c^p\sim50$ T. 
This strongly supports the scenario of the singlet ground state in this compound. 
However, this singlet ground state is much more complicated than that in 3$d$ compounds and should have the internal structure to explain the complicated structures observed in the magnetization and magnetoresistance.

\section*{Acknowledgment}
This work was supported by Grant-in-Aid for Scientific Research on priority Areas gHigh Field Spin Science in 100Th(No.451) from the Ministry of Education, Culture, Sports, Science and Technology(MEXT) and Grant-in-Aid for Young Scientists (B) 22740218 from the Ministry of Education, Science, Sports and Culture.

\end{document}